\documentclass[10pt,conference,twocolumn]{IEEEtran}
\usepackage{wrapfig,booktabs,fancyhdr,amsmath,amsfonts,tabularx,numprint}
\usepackage{cite,bm,bbm,amssymb,amsthm,url,multirow,times,enumitem,comment}
\usepackage{mathtools,siunitx,balance,tikz,adjustbox,graphicx,array}
\usepackage[linesnumbered,ruled]{algorithm2e}
\usepackage[colorlinks=true, linkcolor=black, citecolor=blue, urlcolor=blue]{hyperref}
\usepackage{float}

\usepackage[font=normalsize]{caption}

\usepackage{algpseudocode}

\newcommand{\vb}{\boldsymbol}

\newcommand{\tr}{\mathsf{T}}

\newcommand{\ba}{\begin{array}}
\newcommand{\ea}{\end{array}}

\DeclareMathAlphabet{\mathpzc}{OT1}{pzc}{m}{it}

\begin{document}
	%%% ----------------------------------------------------------Front Matter
	\title{GPU-accelerated Direct Geolocation \\of GNSS Interference}
	\author{\normalsize Jacob S. Clements\IEEEauthorrefmark{1}\IEEEauthorrefmark{2}, and Zachary L. Clements\IEEEauthorrefmark{2}\\
 		\emph{\IEEEauthorrefmark{1}Department of Computer Science, University of Maryland, College Park} \\
		\emph{\IEEEauthorrefmark{2}Department of Aerospace Engineering and Engineering Mechanics, University of Texas at Austin} \\
}
\maketitle

\begin{abstract}

In recent years, there has been a sharp increase in Global Navigation Satellite Systems (GNSS) interference, which has proven to be problematic in GNSS-dependent civilian applications. Many currently deployed GNSS receivers lack the proper countermeasures to defend themselves against interference, prompting the need for alternative defenses.  Satellites in Low Earth Orbit (LEO) provide an opportunity for GNSS interference detection, classification, and localization. The direct geolocation approach has been shown to be well-suited for low SNR regimes and in cases limited to short captures---exactly what is expected for receivers in LEO. Direct geolocation is a single-step search over a geographical grid that enables estimation of the transmitter location directly from correlating raw observed signals. However, a key limitation to this approach is the computational requirements.  This computational burden is compounded for LEO-based receivers as the geographic search space is extensive.  This paper alleviates the computational burden of direct geolocation by exploiting the independence of position-domain correlation across candidate points and time steps: nearly all computation can be accomplished in parallel on a graphics processing unit (GPU). This paper presents and evaluates the performance of GPU-accelerated direct geolocation compared to traditional CPU processing.

\end{abstract}

\newif\ifpreprint
%\preprintfalse
\preprinttrue

\ifpreprint
	
\pagestyle{plain}
\thispagestyle{fancy}  
\fancyhf{} 
\renewcommand{\headrulewidth}{0pt}
\rfoot{\footnotesize \bf University of Maryland Department of Computer Science \\ M.S. Scholarly Paper (\href{https://www.cs.umd.edu/grad/scholarly-papers}{url}), December 2024} \lfoot{\footnotesize \bf Copyright \copyright~2024 by Jacob S. Clements and Zachary L. Clements}

\else

\thispagestyle{empty}
\pagestyle{plain}

\fi

%------------------------------------------------------------------------------------------
\section{Introduction}
Global Navigation Satellite Systems (GNSS) such as the Global Positioning System (GPS) offer precise positioning and timing across the globe. Besides their broad coverage, GNSS are robust to weather conditions and are available to users for free. Due to these useful properties, there are a myriad of technologies that rely on GNSS. For example, positioning information obtained from GNSS is used for navigation in automobiles, ships, and aircraft. Additionally, the timing information provided by GNSS is used to synchronize stock market exchanges and control power grid switches.

Despite their uses, GNSS signals are susceptible to interference due to being extremely weak. For users on Earth, GNSS signals have no more flux density than a 50 W light bulb shining at a distance of 2000 km away \cite{humphreysGNSShandbook}. Because of this, GNSS receivers are highly susceptible to deliberate GNSS interference such as jamming or spoofing. 

Although they are illegal for individuals to use, there are GNSS interference devices on the public market and can be purchased for as little as \$10.  One of such devices is a so-called "personal privacy device" which jams GNSS-based tracking devices so that the tracking device's position cannot be relayed.  Because the authentic GNSS signals are so weak, these personal privacy devices can potentially impact GNSS receivers multiple kilometers away \cite{r_mitch2011cgj}. In 2013, a man in New Jersey was fined \$32,000 when his GPS jammer unintentionally disrupted operations at Newark Airport \cite{cbs_illegal_gps_device}.

Wide-area GNSS interference has been observed near ongoing conflict zones. The civilian maritime and airline industries are encountering GNSS jamming and spoofing at an alarming rate \cite{c4ads2019aboveUsOnlyStars, gebrekidan2023nyt, arraf2024npr, tangel2024wsj, israel_spoofing_dating_apps}. Widespread jamming and spoofing is indicated by monitoring anomalous positioning information broadcast by ships in Automatic Identification System (AIS) messages, and airplanes in Automatic Dependent Surveillance-Broadcast (ADS-B) messages.  All of the civilian GNSS receivers falling victim to GNSS interference are likely unintended targets caught in the electronic warfare crossfire near ongoing conflict zones.

In general, GNSS interference can be separated into two categories: (1) jamming and (2) spoofing.  Jamming is a blunt attack where the jammer transmits strong interference signals at the same frequency as GNSS, with the goal of denial of service.  On the other hand, spoofing is aimed at deception, trying to trick a GPS receiver into inferring a false navigation solution. In the simplest case, jammers can emit a pure tone at the GNSS center frequencies. Chirp jammers can deny a broader range of the GNSS spectrum by varying the frequency of the emitted signal with time to occupy a larger bandwidth. Such chirp jammers appear to be popular in commercially available civil GPS jammers \cite{r_mitch2011cgj, mitch2012kyejam}. More sophisticated attacks such as spoofing can trick GNSS users into computing a false navigation solution by feeding them false GNSS signals \cite{7445815}.

Given the vulnerabilities in GNSS, there has been significant research into navigation robustness and security within the past two decades \cite{psiaki2016gnssSpoofing, jafarnia2012vulReview, humphreysGNSShandbook, psiakiNewBlueBookspoofing, clements2022CpImuSpoofIonItm, clements2023spoofingDetectionGpsWorld, tanil2018insMonitor}. Another significant research thrust has been developing alternate forms of positioning, navigation, and timing (PNT).  Mega-Low Earth Orbit (LEO) constellations can be opportunistically exploited for navigation \cite{komodromos2023IONsimulator, Komodromos2024WeakStarlinkSignal, humphreys2023starlinkSignalStructure, qin2024shortTermFrameTiming, qin2024starlinkTimingProperties, qin2023turret, morgan2024hoocem, humphreys2024gnss}.  Alternate terrestrial radio-based positioning \cite{tenny2022tightTrns, tenny2023robustInsideGnss}, alternate sensors such as cameras \cite{tenny2023XrSlam, catalan2021implementation}, and different receiver architectures  \cite{clements2021bitpackingIonGnss, morrison2024uavDistBeamforming} have been explored. Although such progress has been inspiring, there is still a need to help defend the currently deployed receivers that have not incorporated such defenses.
 
For GNSS users to avoid or account for GNSS interference, it can be helpful to geolocate these interference sources. A solution that achieves this is receivers in LEO, which have been shown to be able to geolocate GNSS interference \cite{lachapelle2021orbitalWardrivingIonGnss, murrian2021leo, clements2023PlansDirectGeo, clements2023PinpointingInsideGnss, clements2022spoofergeo, clements2024DemoSpooferGeo, clementsSpooferSSGJournal}. Satellites in LEO offer worldwide and consistent coverage. LEO satellites are also typically far enough away from GNSS interference sources to acquire and track authentic GNSS signals. The timing information obtained from authentic GNSS signals allows multiple receivers to synchronize samples in their data captures, which allows for precise interference emitter geolocation. With multiple satellites, time difference of arrival (TDOA) and frequency difference of arrival (FDOA) measurements can be exploited to estimate the location of GNSS interference sources. 

The traditional approach to estimating an emitter's location is a two-step approach using time-frequency difference of arrival (T/FDOA) measurements \cite{sidi14_dpf, ho1997geolocation}. In this two-step approach, T/FDOA measurements are first made using the complex ambiguity function (CAF). A CAF is calculated at different snapshots to create a time series of T/FDOA measurements. Then, nonlinear estimation is applied to this time series data to estimate the emitter's location.

This two-step approach has some drawbacks. In the first step, the CAF may have properties that make it hard to extract T/FDOA measurements. Peaks in the CAF are used to determine T/FDOA measurements, but there may be multiple peaks in the CAF if there are multiple emitters or if an emitter is cyclostationary. Because of these ambiguities, a tracking algorithm is usually used to track and associate peaks across CAFs. This tracking and association algorithm struggles in crowded signal environments, especially ones with strong emitters with cyclostationary behavior.

An alternate approach to geolocation using multiple receivers is direct geolocation \cite{reuven2009direct, weiss11_dge, bhatti2015dissertation, clements2023PlansDirectGeo}. In direct geolocation, the TDOA and FDOA are directly parameterized for a single geographical point, given knowledge of the receivers' position and velocity, and then the raw signals correlated in accordance with the T/FDOA.  This is repeated for a set of geographical candidate locations. In the direct approach, cyclostationary behavior of a single emitter does not interfere with the geolocation of other emitters. This makes the direct approach ideal for scenarios with multiple emitters that possibly exhibit cyclostationary behavior and for scenarios that have both strong and weak emitters.

A key challenge these T/FDOA-based geolocation approaches face is computation time. As in the case of the direct approach, data captures that are potentially millions of samples long must be time-aligned according to the calculated TDOA and then FDOA aligned according to the calculated FDOA. Additionally, the direct approach requires this process to be repeated for each candidate point. This computational burden is compounded for LEO-based receivers as the geographic search space is extensive.  For example, for a 10 degree by 10 degree grid defined in Latitude-Longitude-Altitude (LLA) coordinates with 0.01 degree spacing, a million position-domain correlation values must be computed.

This paper attempts to alleviate the computational constraints of direct emitter geolocation by using a graphics processing unit (GPU) to accelerate the computation. The independence of the position-domain correlation values between individual candidate locations is exploited by computing their position-domain correlation values in parallel on a GPU. To account for the vast amount of data being processed on a GPU with limited memory, batch processing is used and then optimized to minimize computational time. A simulation engine and synthetic dataset are developed to show a large speedup in the time required to run direct geolocation compared to traditional CPU processing.

\section{Signal Models}

\subsection{Received Signal Model}
Consider a stationary emitter on the surface of the Earth transmitting a baseband signal $x(t)$ modulating a carrier. When this emitted signal reaches a moving receiver, the observed signal includes both a time delay $\tau$ due to the propagation time and a Doppler shift $f_D$ due to the relative motion. Both $\tau$ and $f_D$ are accurately modeled as constant over a snapshot (short intervals less than 50 ms) as the relative motion over this short interval is negligible. After downconversion, the baseband signal received by the $i$th receiver is given by
\begin{align}
    y_i(t) = x(t-\tau_i)\mathtt{exp}(j2\pi f_{i}t)
\end{align}
where $\tau_i$ and $f_{i}$ are the time delay and Doppler shift for the $i$th receiver. 

Let $N_s$ represent the number of samples, $f_s$ be the sampling rate, and $T_s$ be the time between samples. Let $y_i[k]$ represent the discretized sampled representation of the received signal $y_i(t)$.
\begin{align}
    y_i[k] = y_i(kT_s)
\end{align}
where $k$ goes from 0 to $N_s - 1$. Let the time $t_k$ be the time at the $k$th sample, starting from 0 and going to $T_s(N_s-1)$. 

Consider a stationary emitter with position vector $\vb{p}_\text{e}$ specified in the Earth-centered-Earth-fixed (ECEF) coordinate frame. At a given snapshot, for the $i$th receiver, let $\vb{p}_i$ be its position vector and $\vb{v}_i$ be its velocity vector, both in ECEF. Define the range vector between the emitter and the $i$th receiver $\vb{r}_i$ as
\begin{align}
    \vb{r}_i = \vb{p}_i - \vb{p}_\text{e} 
\end{align}
The range between the emitter and the $i$th receiver $\rho_i$ is then
\begin{align}
    c\tau_i = \rho_i = \sqrt{\vb{r}_i^\tr\vb{r}_i}
\end{align}
where $c$ is the speed of light. The unit range vector $\hat{\vb{r}}_i$ can then be defined as
\begin{align}
    \hat{\vb{r}}_i = \frac{\vb{r}_i}{\rho_i}
\end{align}
The observed Doppler shift at the $i$th receiver $f_i$ is given by
\begin{align}
    f_{i} = -\frac{1}{\lambda} \hat{\vb{r}}_i^\tr \vb{v}_i
\end{align}
where $\lambda$ is the wavelength of the signal.

Different receivers will receive the emitter's signal at different time delays. For two time synchronized receivers, the time difference of arrival between the $i$th receiver and the $j$th receiver is defined as
\begin{align}\label{eqn:TDOA}
    \Delta\tau_{ij} = \tau_j - \tau_i
\end{align}
 The frequency difference of arrival of the $i$th and $j$th receivers is then defined as
\begin{align}\label{eqn:FDOA}
    \Delta f_{ij} &= f_{j} - f_{i}
 \end{align}
These physical received characteristics of a transmitted signal can be exploited to estimate an emitter's position. 

These models assume the relative clock bias and clock drift between the receivers can be compensated for.  Typically, this can be achieved by using the onboard GNSS navigation solution. Each receiver can use their onboard navigation solution to synchronize with GPS time, and in effect, synchronize to the other receivers.

\section{Direct Geolocation}
% \subsection{Two-Step Geolocation}
% In the two-step approach to emitter geolocation, a time series of time and frequency difference of arrival (T/FDOA) measurements is first obtained for each suspected emitter. This is done by computing the CAF at different times during the data capture. The peaks of the CAF are then tracked across time using a multi-target tracking algorithm to produce a time series of T/FDOA measurements for each peak. 

% For each time series of T/FDOA measurements, a nonlinear least-squares estimator is used to find the position of the emitter, which corresponds to the minimum argument of the cost function
% \begin{align}
%     J(\vb{x}) = \frac{1}{2} [z-h(\vb{x})]^TR^{-1}[z-h(\vb{x})]
% \end{align}
% where TODO: either finish explaining two-step or move it to the intro/background

% figure of example cost surface

% discuss drawbacks of two-step approach such as the peak tracking when there are both weak emitters and cyclostationary emitters

% include figure of a CAF with many small and large peaks

% \subsection{Direct Geolocation}
Direct geolocation is a single-step grid-search method that solves directly for the emitter position without the need for intermediate T/FDOA measurements. The CAF is maximized directly by parameterizing the delay and Doppler time histories in terms of the emitter's position, as well as the known receivers' position and velocity.  The cross-correlation between received samples is performed in the position domain instead of the delay-Doppler domain. For a large number of measurements the two-step approach is equivalent to the direct approach \cite{weiss11_dge}.  However, the direct approach outperforms the two-step approach in low SNR regimes and in cases limited to short captures.  This estimator is proven to be maximum likelihood in \cite{weiss11_dge}.

The set of candidate points defined by the user is typically a grid of points defined in LLA coordinates. The user only needs to define these points in latitude and longitude, since the altitude for each point can be constrained to the surface of the Earth, or be retrieved from a digital terrain model.  For the previously stated measurement model to be valid, these coordinates are transformed into an ECEF coordinate system using the standard transformations. Since this coordinate system is rectangular, the measurement model applies if the receiver positions are also expressed in ECEF. 

Let $\vb{s}_i = [\vb{p}_i^\tr, \vb{v}_i^\tr]^\tr$ contain the position and velocity of the $i$th receiver and $\vb{p}_c$ denote the candidate point.  At a snapshot with a fixed receiver geometry, the expected TDOA and FDOA for each $\vb{p}_c$ can be calculated using (\ref{eqn:TDOA}) and (\ref{eqn:FDOA}). For $\vb{p}_c$, let the expected TDOA $\Delta\bar{\tau}_{ij}$ and FDOA $\Delta\bar{f}_{ij}$ be
\begin{align}
    \Delta\bar{\tau}_{ij} &= \mathtt{round}((\Delta\tau_{ij}(\vb{p}_c, \vb{s}_i, \vb{s}_j))/f_s) \\
    \Delta\bar{f}_{ij} &= \Delta f_{ij}(\vb{p}_c, \vb{s}_i, \vb{s}_j)
\end{align}

Then the position-domain correlation value at $\vb{p}_c$ is as
\begin{align}
    S(\vb{p}_c) = \left\lvert\sum_{k=0}^{N_s-1} y_1[k] y_2^*[k+\Delta\bar{\tau}_{ij}] \mathtt{exp}\left(j2\pi \Delta\bar{f}_{ij} t_k\right)\right\lvert
\end{align}

The maximum value of $S$ is the maximum likelihood estimate of the emitter's position \cite{weiss11_dge}.  Direct geolocation is summarized in Algorithm \ref{alg:Directgeolocation}. 
Note, the shape and sharpness of the peak is dependent on: (1) receiver geometry, (2) transmitted waveform, (3) coherent integration time, and (4) received signal power.  As will be shown later, a major advantage of the direct approach is the ability to noncoherently accumulate the position-domain correlation values across snapshots. Due to the transmitted waveform, structures can arise in a position-domain correlation grid that can make detecting peaks in an individual snapshot difficult. When position-domain correlation grids are noncoherently accumulated, those undesirable structures will end up being reduced to noise.  

\begin{figure}[H]
    \centering
    \includegraphics[width=\linewidth]{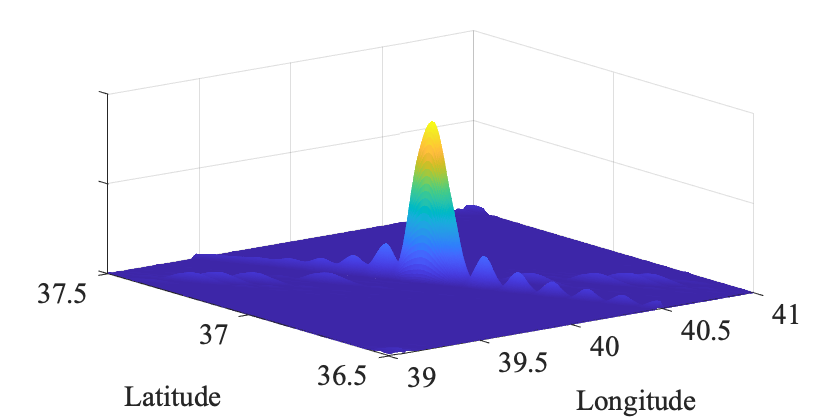}
    \includegraphics[width=\linewidth]{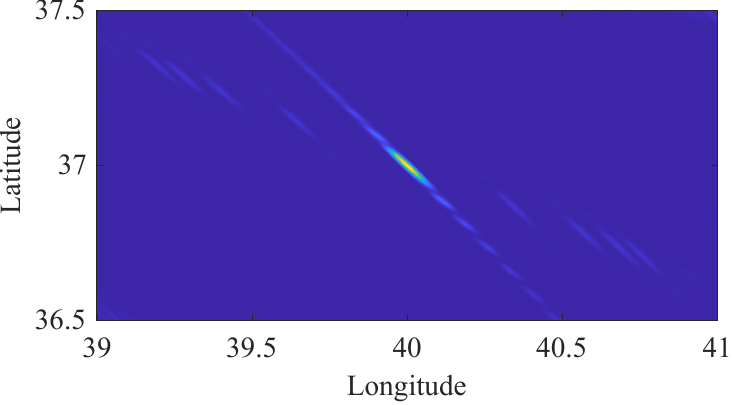}
    \caption{Example position-domain correlation grid from direct geolocation for a single GPS L1 C/A spoofer. }
    \label{fig:exampleDirectGPS}
\end{figure}

A position-domain correlation grid for a GPS L1 C/A spoofer is shown in Fig. \ref{fig:exampleDirectGPS}. As expected, there is a sharp peak that corresponds to the GPS L1 C/A spoofer's position.

\begin{figure}[H]
    \centering
    \includegraphics[width=\linewidth]{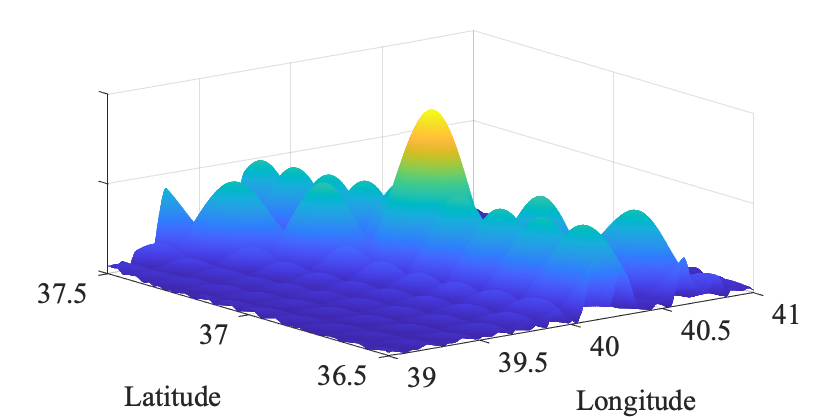}
    \includegraphics[width=\linewidth]{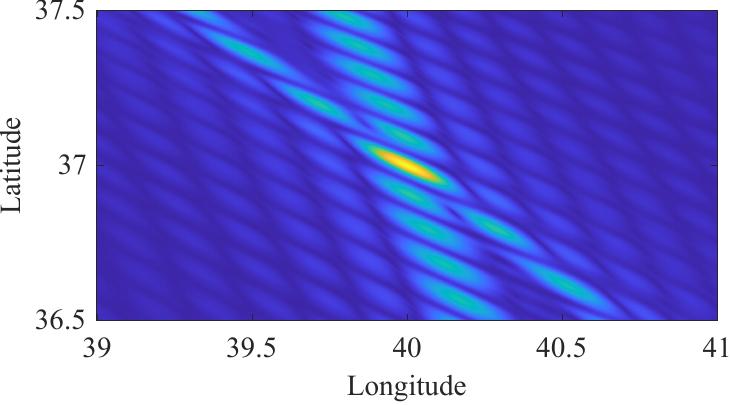}
    \caption{Example position-domain correlation grid from direct geolocation for a sawtooth jammer, which exhibits cyclostationary properties. There are several undesireable sidelobes due to the waveform.}
    \label{fig:exampleDirectSawtooth}
\end{figure}

Shown in Fig. \ref{fig:exampleDirectSawtooth} is an example position-domain correlation grid shown for a sawtooth jammer.  A sawtooth jammer exhibits cyclostationary properties and has a highly structured auto-ambiguity function, creating several sidelobes, even in the position domain.  As will be shown later, this is not problematic when there are multiple snapshots with differing geometries.

\begin{algorithm}
	
	\SetKwInOut{Input}{Input}
	\SetKwInOut{Output}{Output}
	\Input{set of candidate points; raw samples, positions, and velocities from the receivers}
	\vspace{0.25em}
	\Output{Emitter position estimates}
	
	\vspace{1em}
	
	Convert all candidate positions to ECEF
    
        \For{each snapshot}{

            Set $\vb{s}_i = [\vb{p}_i^\tr, \vb{v}_i^\tr]^\tr$

            Set $\vb{s}_j = [\vb{p}_j^\tr, \vb{v}_j^\tr]^\tr$
        
    	\For{each candidate point}{ 
    		Calculate TDOA  $\Delta\tau(\vb{p}_c, \vb{s}_i, \vb{s}_j)$ 
            
                Calculate FDOA $\Delta f(\vb{p}_c, \vb{s}_i, \vb{s}_j)$
            
    		Compute position-domain correlation value $S$
        }
    }
    
    Sum the position-domain correlation grids
    
    Apply a threshold to detect emitters
	\caption{\texttt{Direct Geolocation}}
	\label{alg:Directgeolocation}
\end{algorithm}

\pagebreak

\section{Simulation and Geolocation Results}

This section presents the developed simulation engine and an example capture scenario.  A high-fidelity simulation engine is essential for testing and algorithm development.

\subsection{Simulation Engine}
A high-fidelity simulation engine was developed to explore direct geolocation of GNSS interference sources.  Within this simulation engine framework, the user can specify any arbitrary number of receivers with any geometry.  The user can also specify any number of GNSS interference sources at any location.  Each GNSS interference source transmits one of the four of the most common GNSS interference waveforms: (1) GPS L1 C/A spoofing, (2) pure tone, (3) chirp, and (4) sawtooth.  These transmitted waveforms are generated at baseband and then the received waveform at each receiver is generated.  

The GPS L1 C/A waveform is a spread spectrum signal where the carrier is modulated by a 50 bits per second navigation data sequence, which is further modulated with the C/A code \cite{p_misra06_smp}. Both the navigation data and C/A employ binary phase shift keying (BPSK).  The C/A code is a pseudorandom noise (PRN) sequence that is different for each navigation satellite. Each C/A code is a Gold code with maximal length linear shift register sequence of length 1023, that is upsampled to a chipping rate of 1.023 Mbps and lasts 1 millisecond long.  Every millisecond, the upsampeld 1023 length code repeats.  All of the Gold sequences have good auto-correlation functions, which is important for ranging precision, and uniformly low cross-correlation properties.  GNSS spoofers transmit spoofing signals that mimic the authentic signals, so a would-be GPS L1 C/A spoofer would transmit GPS L1 C/A signals.  Additionally, matched-code and matched-spectrum interference will take on a similar construction.  

A pure tone jammer is by far the simplest form of jamming, as it is simply a tone at the carrier frequency. Chirp and sawtooth jamming waveforms are different, as they can deny a broader range of the GNSS spectrum by varying the frequency of the emitted signal with time to occupy a larger bandwidth. Let $B_\text{chirp}$ be the bandwidth of a chirp, and $T_\text{chirp}$ be the chirp period.  The baseband representation of a chirp $x_\text{chirp}(t)$ is defined as
\begin{align}
    x_\text{chirp}(t) = \mathtt{exp}\left(j2\pi\left( \frac{B_\text{chirp}}{2T_\text{chirp}}t^2 - \frac{B_\text{chirp}}{2}t \right) \right)
\end{align}
A chirp jammer transmits $x_\text{chirp}(t)$ repeatedly.  A sawtooth jammer is slightly different, as it transmits $x_\text{chirp}(t)$ followed by $\mathtt{conj}\left(x_\text{chirp}(t)\right)$ for a period $T_\text{sawtooth} = 2T_\text{chirp}$. 

\begin{figure*}[ht]
    \centering
    \begin{minipage}[b]{0.49\textwidth}
        \includegraphics[width=.9\linewidth]{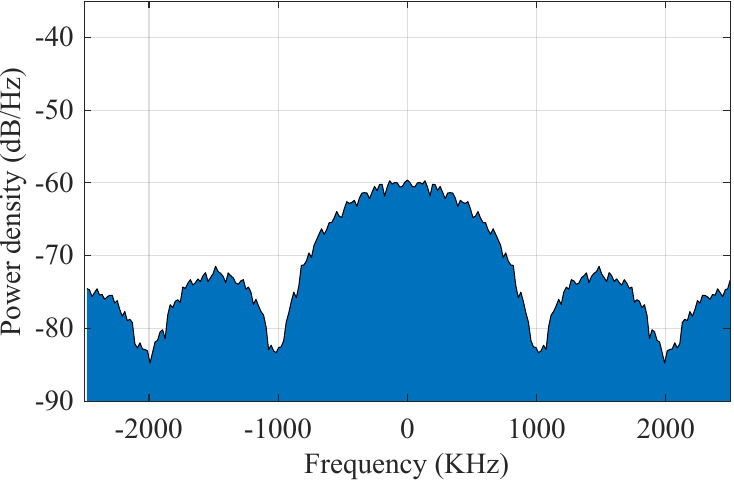} 
        
        \vspace{3mm}
        
        \includegraphics[width=.9\linewidth]{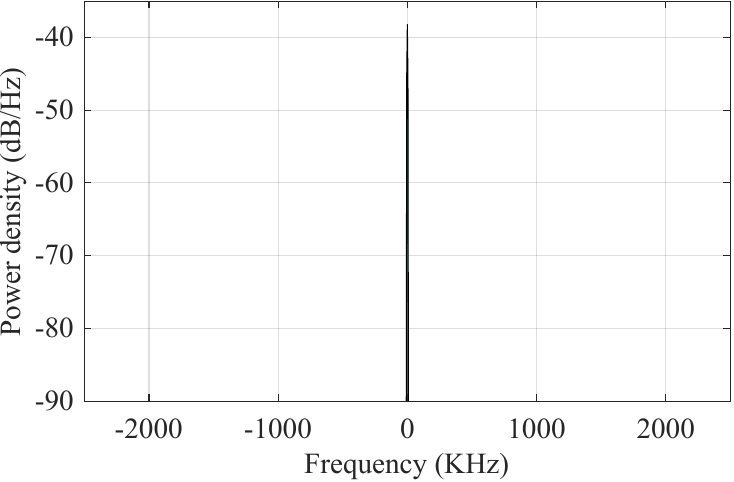}
        
        \vspace{3mm}
        
        \includegraphics[width=.9\linewidth]{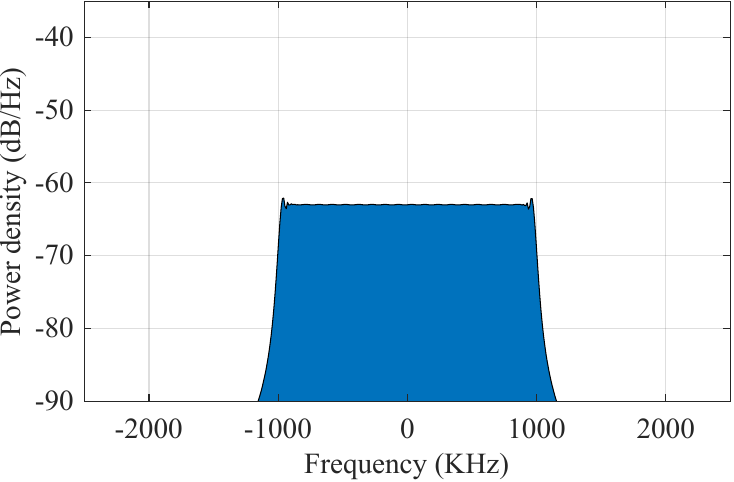}
        
        \vspace{3mm}
        
        \includegraphics[width=.9\linewidth]{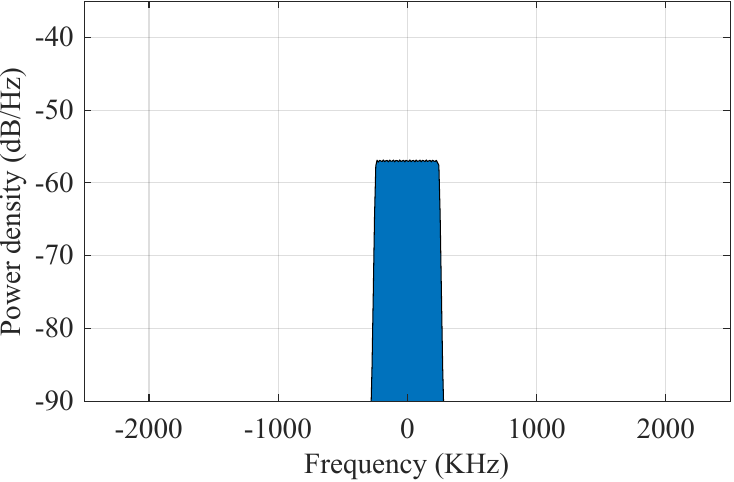}
    \end{minipage}
    \begin{minipage}[b]{0.49\textwidth}
        \includegraphics[width=.92\linewidth]{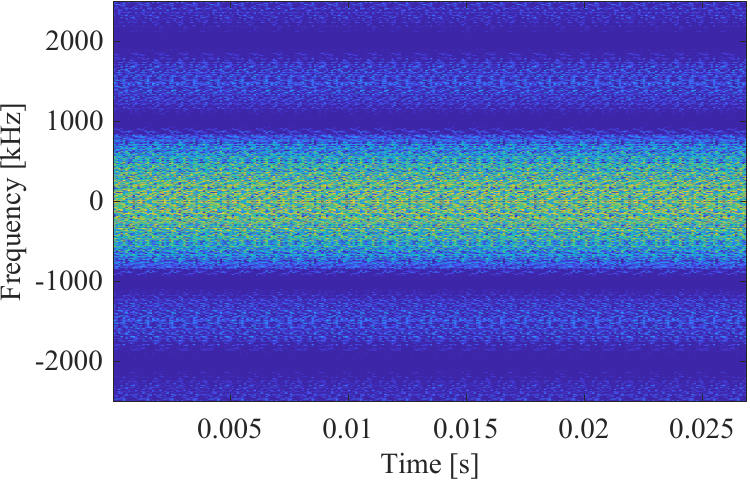}

        \vspace{3mm}
        
        \includegraphics[width=.92\linewidth]{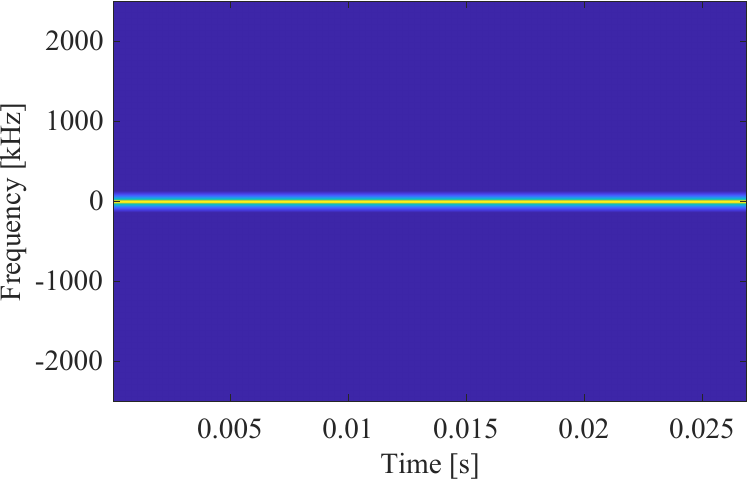}
        
        \vspace{3mm}
                
        \includegraphics[width=.92\linewidth]{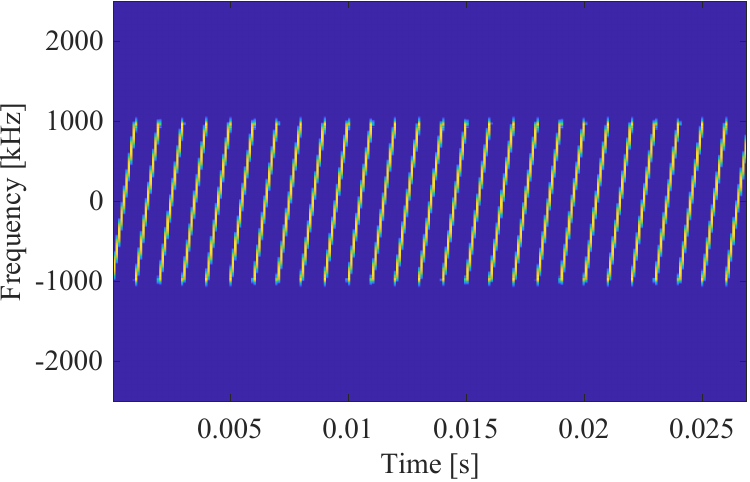}
        
        \vspace{3mm}

        \includegraphics[width=.92\linewidth]{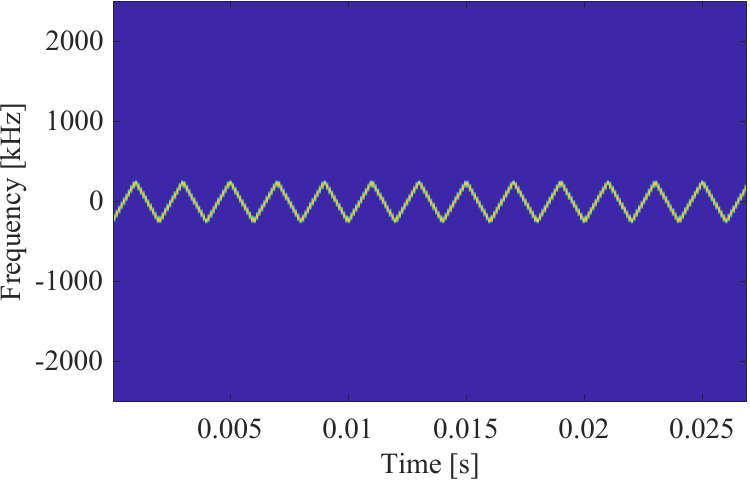}
        
    \end{minipage}
    \caption{Left: Power spectral densities of spoofing, pure tone, chirp, and sawtooth interference waveforms. Right: Spectograms of spoofing, pure tone, chirp, and sawtooth interference waveforms.}
    \label{fig:Waveforms}
\end{figure*}

The power spectral density and spectogram of the GPS L1 C/A, pure tone, chirp, and sawtooth waveforms are shown in Fig. \ref{fig:Waveforms}.

\clearpage

\clearpage

\subsection{Simulated Capture Scenario and Geolocation Results}

\begin{figure}[H]
    \centering
    \includegraphics[width=1\linewidth]{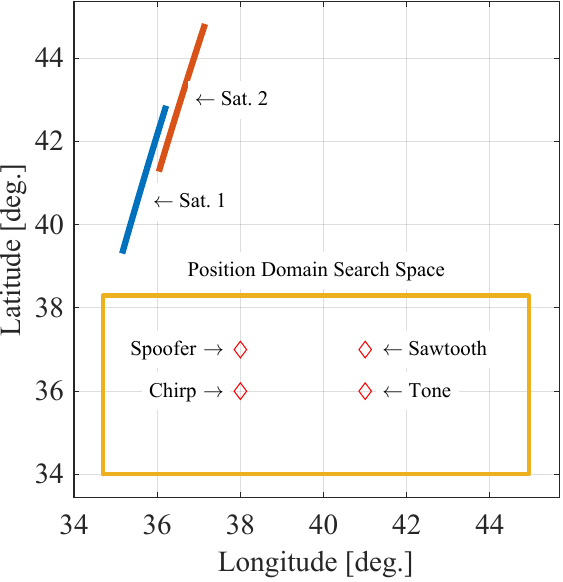}
    \caption{Ground track and search space}
    \label{fig:groundtrack}
\end{figure}

Fig. \ref{fig:groundtrack} shows the relevant physical portions of the simulated capture scenario. The ground tracks of the two LEO-based receivers over 58 seconds are shown.  A GPS L1 C/A spoofer, pure tone jammer, chirp jammer, and sawtooth jammer were simulated to be in a rectangular formation on the surface of the Earth.  The chirp jammer was set to have a bandwidth of 2 MHz and a chirp duration of 20 $\mu$s.  The sawtooth jammer was set to have a bandwidth of 200 kHz and a duration of 5 ms.  All interference sources were set to be transmitting centered at the GPS L1 frequency (1575.42 MHz) at the same power. The receivers performed 58 raw captures, each spaced by a second. Each raw capture was 50 ms long and the sampling rate was set to 5 MHz (complex) centered at the GPS L1 frequency (1575.42 MHz). 

\begin{figure}[H]
    \centering
    \includegraphics[width=1\linewidth]{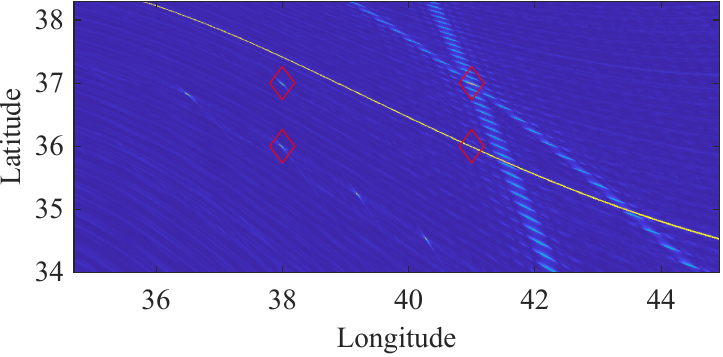}
    \includegraphics[width=1\linewidth]{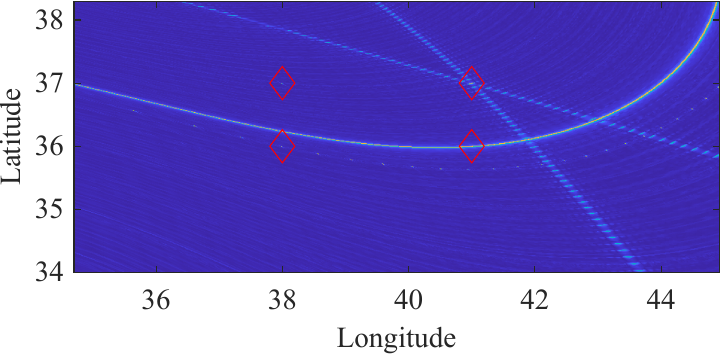}
    \caption{Top: position-domain correlation at the first snapshot. Bottom: position-domain correlation at the final snapshot. }
    \label{fig:pd}
\end{figure}

The position-domain search space shown in Fig. \ref{fig:groundtrack} spans approximately 10$^\circ$ in longitude and approximately 4$^\circ$ in latitude.  The position-domain correlation was computed at each of the 58 snapshots. The position-domain correlation grids for the first snapshot and the last snapshot are shown in Fig. \ref{fig:pd}.  One can immediately see the different correlation structures for the different waveforms. The GPS L1 C/A spoofer creates a single sharp peak, the tone creates a large streak, the chirp jammer creates multiple equally prominent peaks, and the sawtooth jammer creates an ``X" shape.  At a single snapshot, the pure tone and chirp jammers' positions are ambiguous.  But, as seen between the first and last snapshot, the structure of each GNSS interference source changes. This is due to the changing receiver geometry.  However, the peaks at the true emitter positions do not change from the start of the capture compared to the end of the capture.

\begin{figure}[H]
    \centering
    \includegraphics[width=1\linewidth]{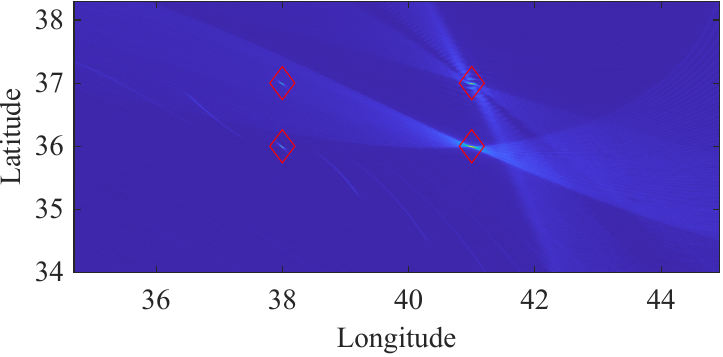}    
    \includegraphics[width=1\linewidth]{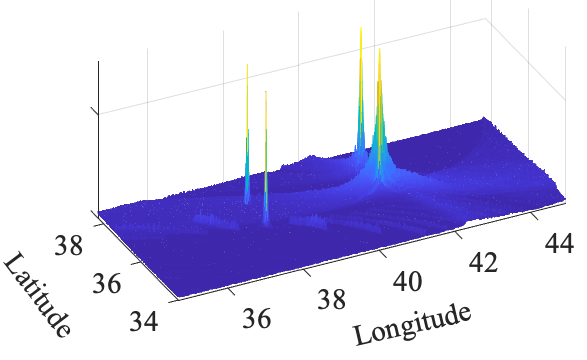}
    \caption{position-domain correlation after accumulating 58 snapshots.  The four distinct sharp peaks represent the MLE of each GNSS interference source's position. }
    \label{fig:pd_all}
\end{figure}

Shown in Fig. \ref{fig:pd_all} is the position-domain correlation grid when the 58 position-domain grids at each snapshot are noncoherent accumulated.  One can immediately appreciate four sharp peaks centered at each of the four GNSS interference sources' true position.  As the receiver geometry changes over the 58 second interval, the positions of the sidelobes and other undesirable structures due to the waveform change.  When position-domain grids are noncoherently accumulated, the undesirable structures are reduced to the noise floor, while the true peaks persist.  This shows how powerful direct geolocation can be in crowded signal environments with waveforms that have undesirable structures.

\section{GPU-Accelerated Analysis}

\subsection{Hardware and Software Setup}
The hardware setup consists of a computer equipped with an Intel 20-core, 28-thread i7-14700K processor and an Nvidia GTX 4060 Ti GPU. In terms of memory, the computer had 32 gigabytes of RAM and the GPU had 16 gigabytes of memory. To utilize the GPU for parallel processing, initial tests were done using Matlab's Parallel Computing Toolbox, PyTorch, and CuPy, with PyTorch being used for the final implementation. PyTorch was also used for the CPU implementation comparisons, and it is worth noting that PyTorch automatically parallelizes many CPU computations under the hood. Unless otherwise noted, all numbers related to performance were averaged over 10 runs.

\subsection{GPU-Accelerated Implementation}
The loops shown in Algorithm \ref{alg:Directgeolocation} are embarrassingly parallelizable since the body of each iteration is independent of all the other iterations for both of the loops. The GPU implementation vectorizes the inner loop to compute the TDOA, FDOA, and position-domain correlation values in parallel on a GPU. Since the memory inside the GPU is limited, this process is batched up. Effectively, the inner for loop becomes a loop over batches of candidate points, and the contents of the for loop (TDOA, FDOA, and position-domain correlation) are all parallelized on the GPU.

Processing each batch of candidate points consists of three steps: (1) load the candidate points and relevant data onto the GPU, (2) perform the computation, and (3) offload the results to CPU memory. To efficiently utilize the limited GPU memory, the final steps of the algorithm, namely the summing over snapshots and thresholding, were left on the CPU. This is done because if the position-domain correlation values are left on the GPU, the amount of free memory on the GPU would deteriorate as the algorithm progresses from one snapshot to the next. Additionally, the summing and thresholding steps were already quite fast and very little performance would be gained by transferring the data back to the GPU to do the computation.

% We noticed that this batching process of loading data and offloading results led to some under utilization of the GPU when it was transferring data around. To address this problem, we used multithreading to schedule data transfers and computation on the GPU in parallel. As shown in Figure (TODO), multithreading increased the overall utilization of the GPU compared to using a single thread. We also simultaneously optimized the number of threads and the batch size for our particular GPU. Figure (TODO) shows the search space we used to optimize the number of threads and our batch size.

\subsection{GPU-Accelerated Analysis}

\begin{figure}[H]
    \centering
    \includegraphics[width=1\linewidth]{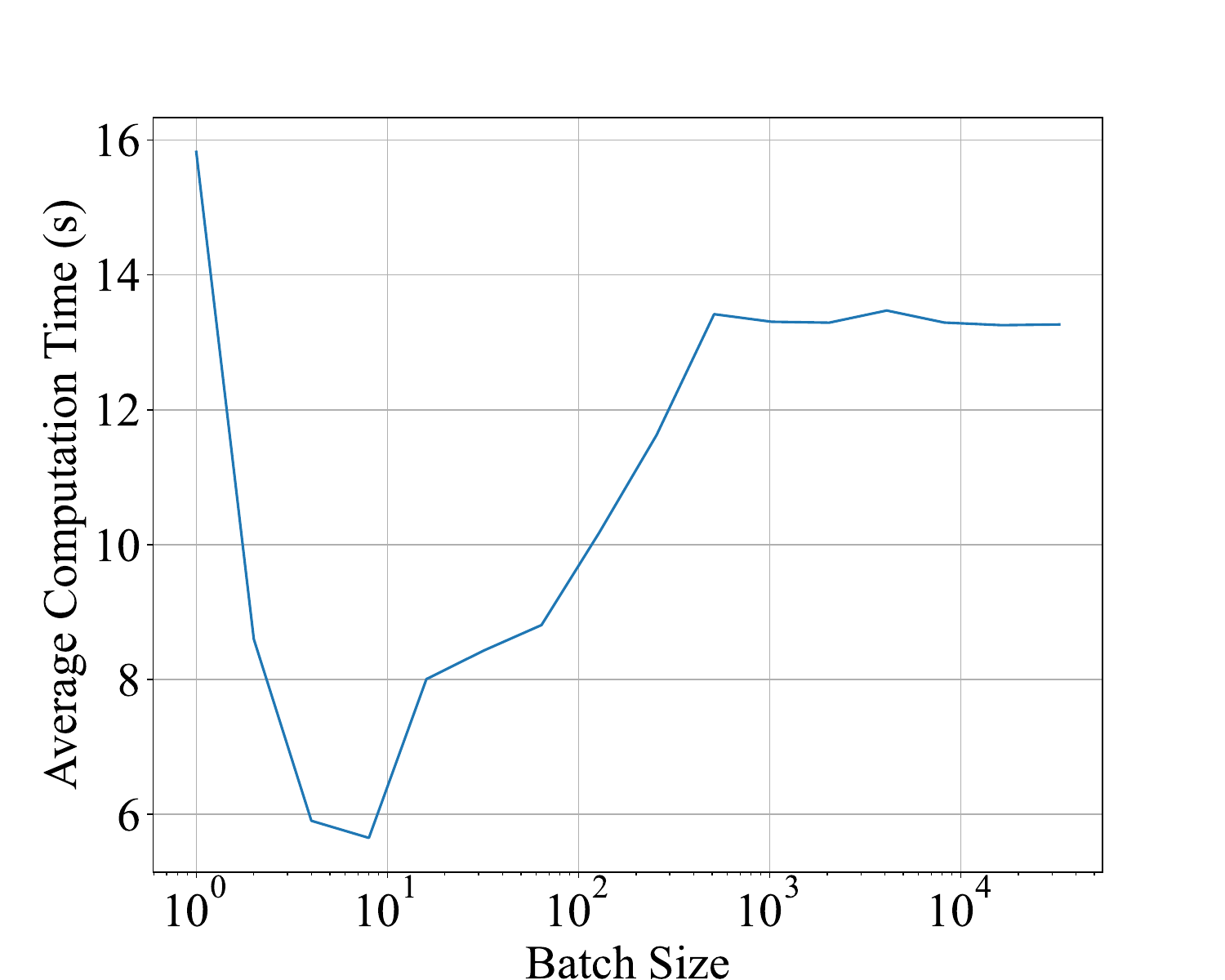}
    \includegraphics[width=1\linewidth]{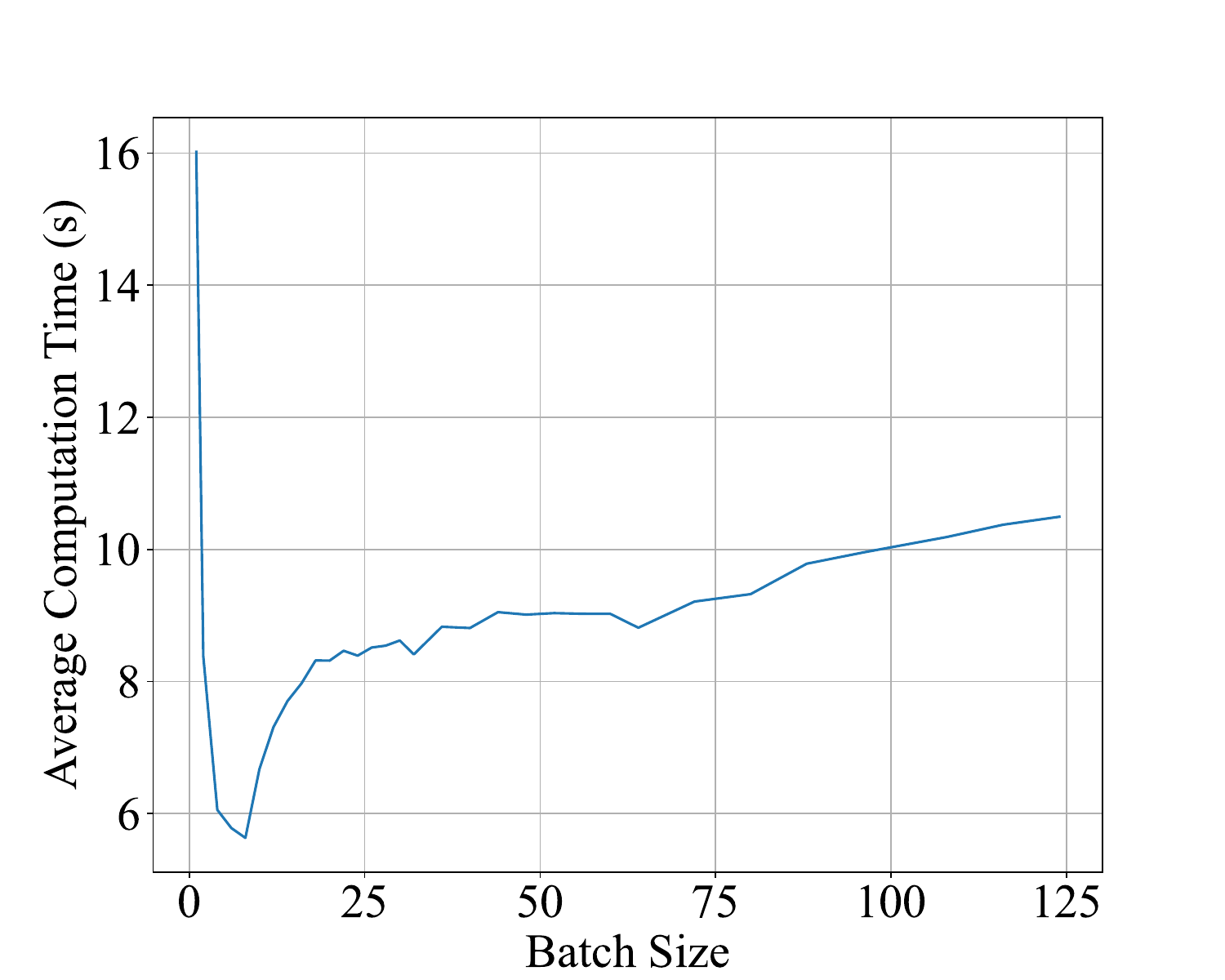}
    \caption{Top: Coarse search of computation time for different batch sizes. The plateau occurs when the GPU memory is full and has to borrow from CPU memory.  Bottom: Fine search of computation time for different batch sizes.}
    \label{fig:batch_size}
\end{figure}

To optimize the batch size, the time required to compute the TDOA, FDOA, and position-domain correlation value for about half a million candidate points is measured. In the coarse search shown in Fig. \ref{fig:batch_size}, there is an immediate drop going from a batch size of 1 to a batch size of 2. After reaching a minimum, the curve rises before plateauing at a batch size of around 500. This plateau happens because the GPU memory is full and has to borrow from CPU memory. After performing a fine search over the region around the minimum in the coarse search, a batch size of 8 was found to be the fastest. Interestingly, batch sizes of 32 and 64 appeared as local minimums, which corresponds with the conventional wisdom that powers of 2 are the most efficient choices for batch sizes. 

\begin{figure}[H]
    \centering
    \includegraphics[width=1\linewidth]{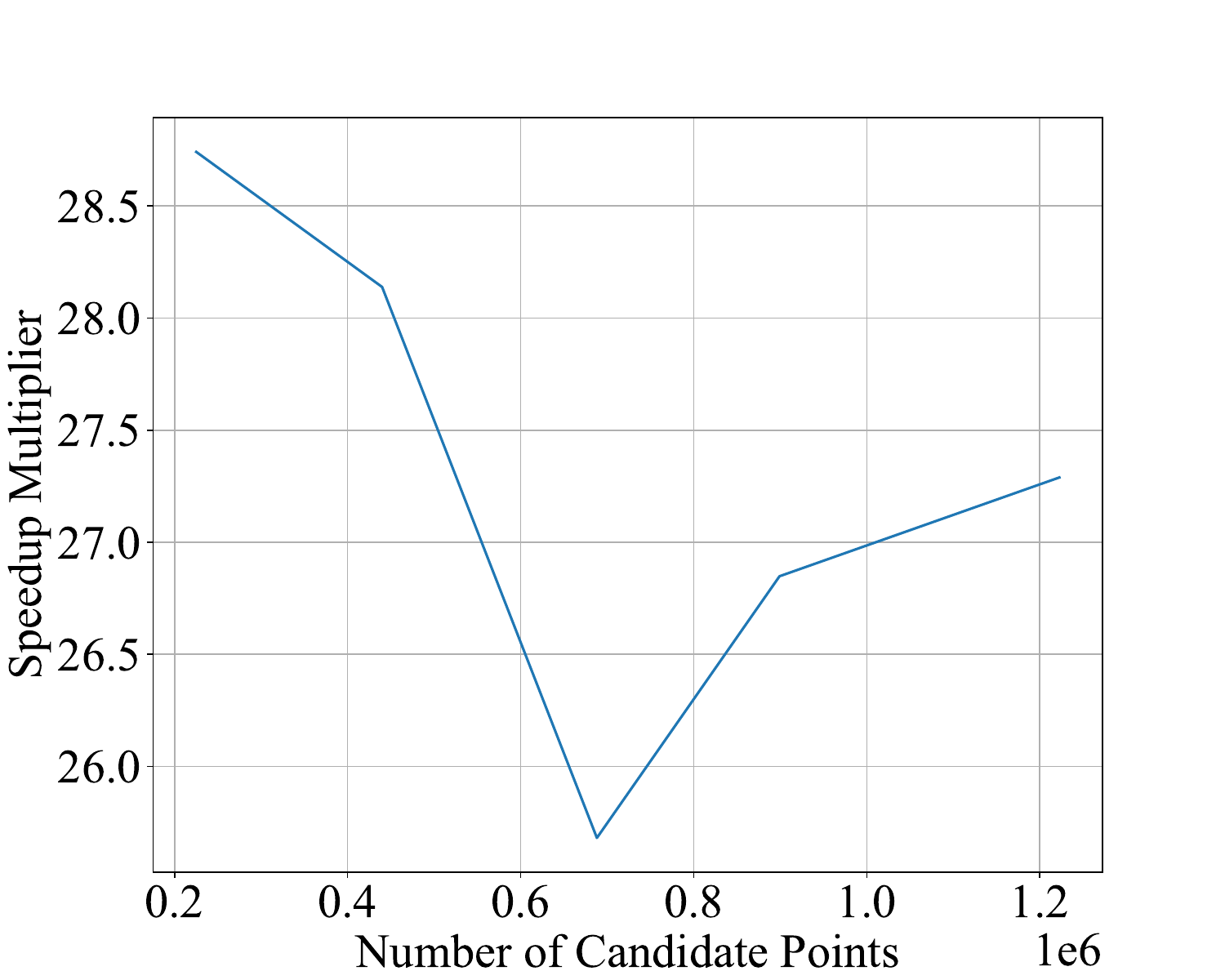}
    \caption{Speedup multiplier when going from CPU to GPU.}
    \label{fig:speedup_number_of_points}
\end{figure}

In general, there were massive improvements in the computational time required to compute the same number of points on a GPU vs on the CPU. Fig. \ref{fig:speedup_number_of_points} illustrates the CPU vs GPU speedup multiplier for a varying number of input candidate points. The speedup remains relatively stable, and Fig. \ref{fig:speedup_number_of_points} shows that the GPU implementation provides a speedup in the range of about 26 times faster to about 28 times faster compared to the CPU implementation.

\section{Conclusion}
GPU-accelerated direct geolocation paired with data from LEO-based receivers is a powerful combination that helps maintain GNSS security. Direct geolocation is a single-step search over a geographical grid that enables estimation of the transmitter location directly from correlating raw observed signals.  However, a key limitation of this approach is the computational time required, since it involves correlating large amounts of data. This computational burden is compounded for LEO-based receivers as the geographic position-domain search space is extensive. This paper addressed this issue by performing the direct geolocation algorithm in parallel on a GPU. Specifically, the TDOA, FDOA, and position-domain correlation values are each computed in parallel. The results show that the GPU-implementation saw speedups of up to around 28 times when compared to a CPU-based implementation.

\bibliographystyle{ieeetr} 
\bibliography{pangea}
\end{document}